\documentclass[aps,preprint]{revtex4}%
\usepackage{amsfonts}
\usepackage{amsmath}
\usepackage{amssymb}
\usepackage{graphicx}%
\begin{document}
\preprint{ }
\title[FBM of HTS]{Fluctuating Cu-O-Cu Bond model of high temperature superconductivity in cuprates}
\author{D. M. Newns and C. C. Tsuei}
\affiliation{IBM T.J. Watson Research Center, Yorktown Heights, NY 10598}
\pacs{PACS number}

\begin{abstract}
\textbf{Twenty years of extensive research has yet to produce a general
consensus on the origin of high temperature superconductivity (HTS). However,
several generic characteristics of the cuprate superconductors have emerged as
the essential ingredients of and/or constraints on any viable microscopic
model of HTS. Besides a }$T_{c}$ \textbf{of order }$100%
%TCIMACRO{\unit{K}}%
%BeginExpansion
\operatorname{K}%
%EndExpansion
$\textbf{, the most prominent on the list include a d-wave superconducting gap
with Fermi liquid nodal excitations, a d-wave pseudogap\ with the
characteristic temperature scale }$T^{\ast}$\textbf{, an} \textbf{anomalous
doping-dependent oxygen isotope shift, nanometer-scale gap inhomogeneity,
etc.. The key role of planar oxygen vibrations implied by the isotope shift
and other evidence, in the context of CuO}$_{2}$ \textbf{plane symmetry and
charge constraints from the strong intra-3}$d$\textbf{\ Coulomb repulsion }%
$U$\textbf{, enforces an anharmonic mechanism in which the oxygen vibrational
amplitude modulates the strength of the in-plane Cu-Cu bond. We show, within a
Fermi liquid framework, that this mechanism can lead to strong d-wave pairing
and to a natural explanation of the salient features of HTS.}

\end{abstract}
\date[June 8 2006]{}
\startpage{1}
\endpage{1}
\maketitle

Great strides towards understanding HTS have been made during the last twenty
years of intense experimental and theoretical research on cuprate
superconductors \cite{Leggett}, \cite{Bonn}. However, there is no general
consensus on the origin of HTS. Especially there is no report on a specific
microscopic pairing mechanism that is capable of encompassing the complex
phenomenology of the superconducting and normal states consistently in one
theoretical model.

Among the highly unconventional properties of the cuprate superconductors are
the $d$-wave symmetry of the superconducting gap \cite{Tsuei} and the presence
of a pseudogap \textit{also} with $d$-wave symmetry \cite{Norman}. Theoretical
insight is provided by study of the low-energy excitations around the node in
the $d$-wave gap, where disparate low-temperature experiments including
specific heat \cite{KAM}, transport \cite{PALee}, \cite{Taillefer},
\cite{Chiao}, \cite{Proust}, \cite{Sutherland}, \cite{Bel}, and penetration
depth \cite{Hardy}, together with angle-resolved photoemission spectroscopy
(ARPES) \cite{ZX} can all be interpreted in terms of a Fermi liquid
description. This suggests the plausibility of a broadly BCS framework
\cite{Bonn}, of which an important consequence is that the large on-site
Coulomb repulsion $U$ \cite{PWA} does not enter into $d$-wave pairing to first order.

In conventional superconductivity the isotope exponent $\alpha$ has been key
in signalling the role of phonons in the pairing mechanism. HTS exhibits a
universal, anomalous doping-dependent isotope shift \cite{alpha1},
\cite{Tallon} which shows that phonons are again playing a key role, but in
some unconventional manner, and any viable microscopic model must be able to
account for this. A further challenge to theory is the recently observed
spatial inhomogeneity of the gap at the nanometer scale \cite{McElroy}.

Recently interest in the role of electron-phonon coupling has been rekindled
by mounting experimental support, which, in addition to the isotope shift,
includes electronic Raman scattering, \cite{Opel}, \cite{Gallais}, ARPES
\cite{Cuk1}, \cite{Cuk2}, \cite{Gweon}, \cite{Zhou}, inelastic neutron
scattering \cite{Pintschovius}, x-ray absorption fine structure \cite{Saini},
low-temperature STM/STS \cite{Jinho}, and isotope effect in penetration depth
measurements \cite{Khasanov}. Extensive theoretical studies \cite{Bussman1},
\cite{Bussman2}, \cite{Muller} \cite{Marvin}, \cite{Kulic}, \cite{Cappelluti},
\cite{Scalapino}, \cite{Cuk1}, \cite{DHLee}, \cite{Schuttler} have shed light
on the relevance of electron-phonon interaction in understanding HTS.

Here, in re-examining phonon coupling, we find a novel microscopic pairing
mechanism which can indeed explain the essence of HTS phenomena: high $T_{c}%
$'s, $d$-wave pairing, $d$-symmetry pseudogap, anomalous isotope shift, and
nanoscale gap inhomogeneity. Although the effects of $U$ are manifested
\cite{PWA}, here we shall focus on interactions directly responsible for
pairing, allowing for $U$ \textit{indirectly} by decoupling from on-site
charge fluctuations. We start by considering how oxygen vibrations can modify
electronic motion in the CuO$_{2}$ plane, the universal active component of
HTS materials. The CuO$_{2}$ plane is a square lattice of divalent Cu ions
with oxygen ions located at the centers of the Cu-Cu bond (Fig. 1a). Only the
highest-lying 3$d_{x^{2}-y^{2}}$ orbital (Fig. 1b) plays an active role,
crystal field effects demoting the other Cu 3$d^{9}$ orbitals to core-like
status.%
%TCIMACRO{\FRAME{ftbpFU}{5.046in}{3.6437in}{0pt}{\Qcb{(a) The unit cell in the
%CuO$_{2}$ plane, with Cu atoms (yellow) and O atoms (red), showing $x$, $y$,
%and $z$ vibrational modes (arrows). (b) The Cu 3$d_{x^{2}-y^{2}}$ and O
%2$p_{x}$ orbitals, illustrating effect of an O $z$-displacement (green
%arrows), positive sense (top panel), and negative sense (bottom panel).}%
%}{\Qlb{Fig. 1}}{fig_1ab3.eps}{\special{ language "Scientific Word";
%type "GRAPHIC";  display "USEDEF";  valid_file "F";  width 5.046in;
%height 3.6437in;  depth 0pt;  original-width 10.1434in;
%original-height 6.8822in;  cropleft "0";  croptop "1";  cropright "0.8903";
%cropbottom "0.1660";  filename 'Fig_1ab3.eps';file-properties "XNPEU";}}}%
%BeginExpansion
\begin{figure}
[ptb]
\begin{center}
\includegraphics[
trim=0.000000in 1.142445in 1.112731in 0.000000in,
height=3.6437in,
width=5.046in
]%
{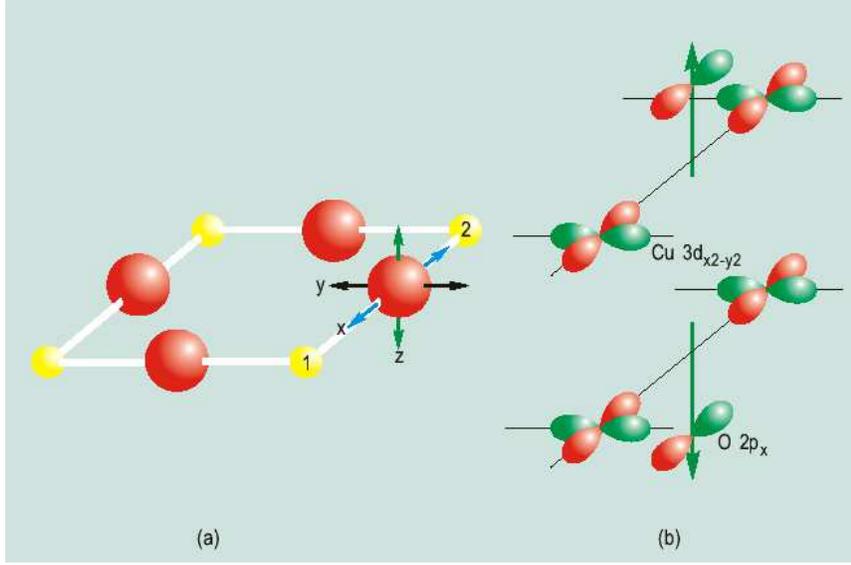}%
\caption{(a) The unit cell in the CuO$_{2}$ plane, with Cu atoms (yellow) and
O atoms (red), showing $x$, $y$, and $z$ vibrational modes (arrows). (b) The
Cu 3$d_{x^{2}-y^{2}}$ and O 2$p_{x}$ orbitals, illustrating effect of an O
$z$-displacement (green arrows), positive sense (top panel), and negative
sense (bottom panel).}%
\label{Fig. 1}%
\end{center}
\end{figure}
%EndExpansion

Let us look at a single Cu-Cu bond between 3$d_{x^{2}-y^{2}}$ orbitals located
on atoms labeled 1 and 2 (Fig. 1). The kinetic energy associated with the
unperturbed direct bond 1-2 comes, within a one-band model, from the
nearest-neighbor hopping matrix element $t$%
\begin{equation}
h_{12}^{e}=-t\sum_{\sigma}\left(  c_{1,\sigma}^{+}c_{2,\sigma}+c_{2,\sigma
}^{+}c_{1,\sigma}\right)  , \label{he}%
\end{equation}
where the $c_{i,\sigma}^{+}$ ($c_{i,\sigma}$) are Fermion creation
(destruction) operators for Cu site $i$ and spin $\sigma$. The intersite
hopping actually occurs via superexchange, i.e. wavefunction overlap between
the 3$d_{x^{2}-y^{2}}$ orbital of Cu1, the intrabond oxygen 2$p_{x}$
longitudinal with the bond, and the 3$d_{x^{2}-y^{2}}$ orbital of Cu2 (Fig.
1b), sensitizing it to the local oxygen vibrational degrees of freedom\ (Fig.
1a). Consider the effect, sketched in Fig. 1b, of, for example, the
out-of-plane oxygen displacement $z$, which is to reduce the overlap between
both 3$d_{x^{2}-y^{2}}$ orbitals and the oxygen 2$p_{x}$ orbital, thus
reducing the effective coupling $t$. But due to the local centrosymmetry of
the O-site, this $t$-reduction effect is the same irrespective of the sign of
the displacement $z$, so that its lowest order expression is as $z^{2}$. Hence
the electron-vibrator term in the Hamiltonian must have the unusual second
order form of coupling
\begin{equation}
h_{12}^{ev}=\frac{v}{2}z^{2}\sum_{\sigma}\left(  c_{1,\sigma}^{+}c_{2,\sigma
}+c_{2,\sigma}^{+}c_{1,\sigma}\right)  ,
\end{equation}
where $v$ is the coupling strength. This electron-vibrator coupling causes the
Cu-Cu bond strength $t$ to \textit{fluctuate} with the oxygen vibrator square
amplitude $z^{2}$- hence the description Fluctuating Bond Model (FBM). The two
other oxygen vibrational modes, $x$ and $y$, can also couple to the bond
strength in a similar manner. In addition to coupling to the bond, the
vibrational mode $x$, longitudinal to the bond, can also displace charge onto
the Cu1 and Cu2 sites; however since charge accumulation on the Cu sites is
resisted by the large intrasite Coulomb interaction $U$, we do not believe
that on-site charge displacement can be important, and ignore it in our model
at the present maximally simplified stage.%
%TCIMACRO{\FRAME{ftbpFU}{5.7556in}{3.2002in}{0pt}{\Qcb{Left (a): Bare oxygen
%anharmonic potential (3) full (dashed) curves with positive (negative)
%harmonic coefficient $m\omega_{0}^{2}$. Right (b): Binding of two
%quasiparticles (e.g. holes h) in a Cu-Cu bond, as given by Eq. (5).}%
%}{\Qlb{Fig2}}{fig_1cd.eps}{\special{ language "Scientific Word";
%type "GRAPHIC";  display "USEDEF";  valid_file "F";  width 5.7556in;
%height 3.2002in;  depth 0pt;  original-width 9.6548in;
%original-height 5.7519in;  cropleft "0";  croptop "0.9758";
%cropright "0.9901";  cropbottom "0.3722";
%filename 'Fig_1cd.eps';file-properties "XNPEU";}}}%
%BeginExpansion
\begin{figure}
[ptb]
\begin{center}
\includegraphics[
trim=0.000000in 2.140857in 0.095582in 0.139196in,
height=3.2002in,
width=5.7556in
]%
{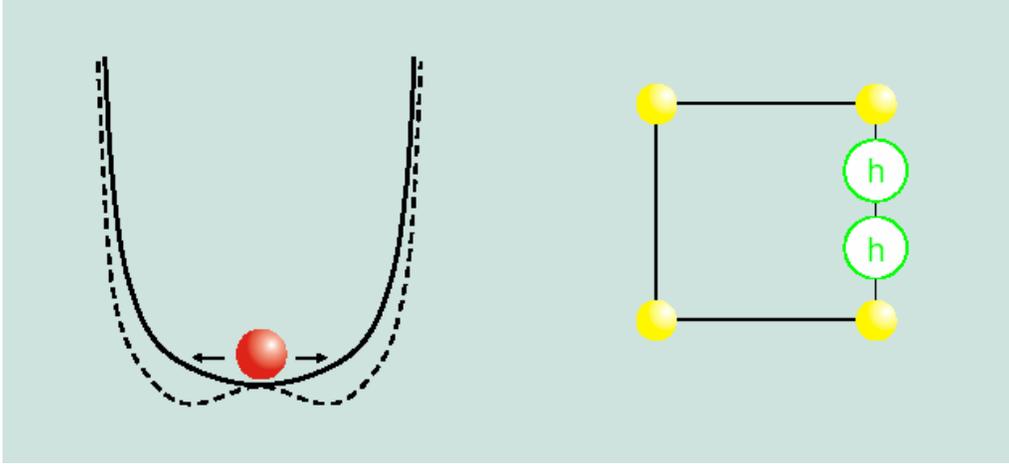}%
\caption{Left (a): Bare oxygen anharmonic potential (3) full (dashed) curves
with positive (negative) harmonic coefficient $m\omega_{0}^{2}$. Right (b):
Binding of two quasiparticles (e.g. holes h) in a Cu-Cu bond, as given by Eq.
(5).}%
\label{Fig2}%
\end{center}
\end{figure}
%EndExpansion

The oxygen vibrational degree of freedom needs to embody an anharmonic
potential term in a theory with nonlinear electron-vibrator coupling, e.g. for
the $z$-mode the vibrator Hamiltonian\ should have the form%
\begin{equation}
h_{12}^{v}=\frac{p_{z}^{2}}{2m}+\frac{1}{2}m\omega_{0}^{2}z^{2}+\frac{w}%
{8}z^{4},
\end{equation}
where a fourth-order potential with coefficient $w$ (again, centrosymmetry is
used to eliminate a cubic term) has been included, in addition to the
conventional harmonic terms in which $p_{z}$ is oxygen momentum, $m$ effective
mass, and $\omega_{0}$ the harmonic vibrator frequency. The potential may be a
flattened well or a double well, depending on the sign of $\omega_{0}^{2}$
(see Fig. 2a). A quartic anharmonic potential is not unique in perovskite
physics, a similar potential plays a key role in the theory of
ferroelectricity \cite{Thomas}.

In the following the FBM Pairing Interaction will be derived systematically.
First, we give a non-rigorous argument leading to a physical picture of the
pairing by completing the square in the highest-order interaction. Introducing
the compact notation $X_{12}$ as the bond order operator, describing the
strength of the Cu1-Cu2 electronic coupling,
\begin{equation}
X_{12}=\sum_{\sigma}\left(  c_{1,\sigma}^{+}c_{2,\sigma}+c_{2,\sigma}%
^{+}c_{1,\sigma}\right)  ,
\end{equation}
and completing the square (classically valid in the Fig 2a double well case
when $-m\omega_{0}^{2}>>v\left\langle X_{12}\right\rangle $)
\begin{equation}
\frac{w}{8}z^{4}+\frac{vz^{2}}{2}X_{12}=\frac{w}{8}\left(  z^{2}+\frac{2v}%
{w}X_{12}\right)  ^{2}-\frac{K}{2}X_{12}^{2},
\end{equation}
we find a new, \textbf{attractive} electron-electron (or hole-hole)
interaction in the intra-bond channel, of strength $K$, where $K=v^{2}/w$. $K$
is the key interaction in the FBM. Because the interaction $K$ acts via the
bond operator $X_{12}$, it is seen from the form $X_{12}^{2}$ to lie in the
intra-bond channel (see Fig. 2b) as opposed to being on-site, and with the
$y$-bond included the Fourier transform of the interaction leads to the d-wave
factor $\cos q_{x}-\cos q_{y}$, resulting in pairing and pseudogap phenomena
of \textit{d-wave symmetry}. There is experimental evidence for linking the
HTS pairing mechanism with the oxygen $z$-vibrational mode: if the vibrator is
localized in the $z$-direction by large static CuO$_{2}$ plane buckling
\cite{buckling},$\ T_{c}$ is reduced and can go to zero. Hence a key role in
pairing must indeed be played by the oxygen degree of freedom.

We are now ready to write down the complete FBM Hamiltonian, as a sum of
electronic, vibrator, and coupling terms:
\begin{equation}
H^{FBM}=H^{e}+H^{v}+H^{ev}. \label{FBM}%
\end{equation}
Here the electronic term includes hopping over longer ranges than the nearest
neighbor hopping considered in Eq(\ref{he}):
\begin{equation}
H^{e}=-\frac{1}{2}\sum_{\mathbf{i,j},\sigma}t\left(  \mathbf{i}-\mathbf{j}%
\right)  c_{\mathbf{i},\sigma}^{+}c_{\mathbf{j},\sigma}, \label{he2}%
\end{equation}
where $\mathbf{i}$ denotes the 3$d_{x^{2}-y^{2}}$ orbital on lattice site
$\mathbf{i=(}i_{x},i_{y})$ in the 2D square lattice of Cu ions. The strongest
interaction is the nearest neighbor hopping integral $t(\pm1,0)=t(0,\pm1)=t$,
followed by the next-nearest neighbor interaction $t(\pm1,\pm1)=t^{\prime}$,
and then the 3rd-nearest neighbor interaction $t(\pm2,0)=t(0,\pm
2)=t^{\prime\prime}$. The band eigenvalues $\epsilon_{\mathbf{k}}$ of
(\ref{he2}) are $\epsilon_{\mathbf{k}}=$ $-2t(\cos k_{x}+\cos k_{y})$
$-4t^{\prime}\cos k_{x}\cos k_{y}$ $-2t^{\prime\prime}(\cos2k_{x}+\cos2k_{y})$
in units where lattice constant$=1$.

In the vibrational term
\begin{equation}
H^{v}=\sum_{\mathbf{i,\alpha}}\left[  \frac{1}{2m}p_{\mathbf{i},\alpha}%
^{2}+\frac{m\omega_{0}^{2}}{2}x_{\mathbf{i},\alpha}^{2}+\frac{w}{8n}\left(
x_{\mathbf{i},\alpha}^{2}\right)  ^{2}\right]  ,
\end{equation}
the bonds are relabelled for greater notational convenience, in terms of a Cu
site $\mathbf{i}$ and a direction $\alpha=x$ or $y$ away from $\mathbf{i}$ in
the positive axis direction. The vibrational modes are approximated as local
(Einstein) and isotropic. The notation $p^{2}$, $x^{2}$ implies $\sum_{s}%
p_{s}^{2}$, $\sum_{s}x_{s}^{2}$ where $s$ is polarization ($s=$ transverse to
plane, longitudinal to bond, or in plane transverse to bond). The mode
degeneracy is $n$.

The bond order operators are defined using the same bond notation
\begin{equation}
X_{\mathbf{i,\alpha}}=\sum_{\sigma}\left[  c_{\mathbf{i},\sigma}%
^{+}c_{\mathbf{i+}\widehat{\mathbf{x}}_{\alpha},\sigma}+c_{\mathbf{i+}%
\widehat{\mathbf{x}}_{\alpha},\sigma}^{+}c_{\mathbf{i},\sigma}\right]  ,
\end{equation}
where $\widehat{\mathbf{x}}_{\alpha}$ is a unit vector along the direction
$\alpha$. In terms of these, the coupling Hamiltonian is%
\begin{equation}
H^{ev}=\frac{v}{2\sqrt{nn_{s}}}\sum_{\mathbf{i,\alpha}}x_{\mathbf{i,\alpha}%
}^{2}X_{\mathbf{i,\alpha}}. \label{FBMev}%
\end{equation}
The prefactor includes a spin degeneracy $n_{s}$.

To solve the FBM we shall in this paper use the standard weak-coupling
approach, based on an electron gas as the unperturbed system. To develop a
perturbation expansion in the absence of the Migdal theorem in HTS
\cite{Tsuei}, we adopt the $1/N$ expansion technique, where $N$ is degeneracy,
e.g. orbital or spin degeneracy. The $1/N$ expansion works well e.g. for the
Kondo problem \cite{Read-Newns}, the results remaining physical down to spin
degeneracy $N=2$. Here we systematically co-expand in the inverse of the mode
degeneracy $n$ and the spin degeneracy $n_{s}$ using a path integral approach
\cite{Coleman} (see supplementary material), meaning by expressions such as
"$1/N$" the joint orders \thinspace$1/n$ and $1/n_{s}$.%
%TCIMACRO{\FRAME{ftbpFU}{4.5446in}{2.9603in}{0pt}{\Qcb{(a) The interaction
%between two pairing quasiparticles $\mathbf{k}$ and $\mathbf{-k}$, exchanging
%momentum $\mathbf{q}$ and frequency $\omega_{n}$, is a product of $d$-wave
%form factors $\xi_{\mathbf{k,k+q}}^{2}$ (yellow circles) and potential
%propagator $V\left(  \mathbf{q},\omega_{n}\right)  $(b) First approximation to
%$V$ based on $v^{2}\times$ two-boson propagator. Dashed lines represent single
%vibrator (boson) propagator, red circles electron-boson interaction $v$. (c)
%The leading-$N$ self-energy corrections to $V$ (see analogous corrections in
%the heavy fermion problem \cite{Lavagna}) come from boson-boson interaction
%$w$ and $v^{2}\times$ Response Function $R_{dd}$. Full lines represent fermion
%quasiparticle propagator. Green square represents boson-boson interaction $w$,
%fermion lens is response Function $R_{dd}$. }}{}{fig3_newy.eps}%
%width 4.5446in;  height 2.9603in;  depth 0pt;  original-width 10.6415in;
%original-height 8.1327in;  cropleft "0";  croptop "0.9084";
%cropright "0.8998";  cropbottom "0.1439";
%filename 'Fig3_newy.eps';file-properties "XNPEU";}}}%
%BeginExpansion
\begin{figure}
[ptb]
\begin{center}
\includegraphics[
trim=0.000000in 1.170296in 1.066278in 0.744955in,
height=2.9603in,
width=4.5446in
]%
{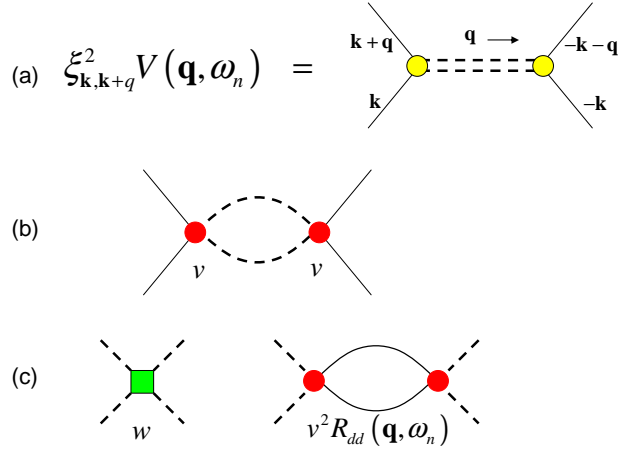}%
\caption{(a) The interaction between two pairing quasiparticles $\mathbf{k}$
and $\mathbf{-k}$, exchanging momentum $\mathbf{q}$ and frequency $\omega_{n}%
$, is a product of $d$-wave form factors $\xi_{\mathbf{k,k+q}}^{2}$ (yellow
circles) and potential propagator $V\left(  \mathbf{q},\omega_{n}\right)  $(b)
First approximation to $V$ based on $v^{2}\times$ two-boson propagator. Dashed
lines represent single vibrator (boson) propagator, red circles electron-boson
interaction $v$. (c) The leading-$N$ self-energy corrections to $V$ (see
analogous corrections in the heavy fermion problem \cite{Lavagna}) come from
boson-boson interaction $w$ and $v^{2}\times$ Response Function $R_{dd}$. Full
lines represent fermion quasiparticle propagator. Green square represents
boson-boson interaction $w$, fermion lens is response Function $R_{dd}$. }%
\end{center}
\end{figure}
%EndExpansion

The interaction which scatters a pair $(\mathbf{k,-k})$ to $(\mathbf{k+q,-k-q}%
)$ (Fig. 3) differs from the usual single phonon propagator structure in BCS,
because there is no single electron-phonon interaction term in the FBM Eq's
(\ref{FBM}-\ref{FBMev}). The structure is instead a \textit{two-boson
}propagator (Fig. 3b) with self-energy insertions, the leading-$N$ self energy
involving the boson-boson interaction $w$ and $v^{2}\times$ a fermion response
function (Fig. 3c). Because the response function can go from bond $\alpha=x$
or $y$ to bond $\beta=x$ or $y$ it is a $2\times2$ matrix. The matrix can be
simplified by noticing the dramatic \textit{divergence} produced by the
self-energy insertions at small transferred wavevector $q$. This small-$q$
divergence comes from fluctuations having $d$-wave symmetry, the $s$-wave part
being nonsingular and uninteresting. Therefore, our simplifying procedure is
to reduce the $2\times2$ propagator matrix to a scalar by projecting out
$s$-symmetry charge fluctuations and retaining only the $d$-symmetry ones, the
projection procedure being\ implemented in path integral formalism (see
supplementary material). The vibrator amplitude/$X$-operator fluctuations
around each Cu site now form a $d_{x^{2}-y^{2}}$-like pattern (see Fig. 4).
This selection of the $d$-channel has the additional physical merit that
charge flow into and out of each Cu site is balanced (Fig. 4), i.e. there is
zero net site charge accumulation, compatible with the accepted large Coulomb
repulsion $U$ on each site, which inhibits such charge fluctuations.%
%TCIMACRO{\FRAME{ftbpFU}{5.5578in}{2.7523in}{0pt}{\Qcb{The Fluctuating Bond
%field after projecting out $s$-fluctuations has $d$-symmetry, with (left
%panel) O-amplitude (dumbells) large in $y$-bonds, small in $x$-bonds, the
%reverse in right panel. Panels also illustrate opposite phases of the static
%$d$-wave CDW (18).}}{\Qlb{Fig4}}{fig_3.eps}%
%width 5.5578in;  height 2.7523in;  depth 0pt;  original-width 11.0056in;
%original-height 8.4968in;  cropleft "0";  croptop "0.9758";  cropright "1";
%cropbottom "0.3376";  filename 'Fig_3.eps';file-properties "XNPEU";}}}%
%BeginExpansion
\begin{figure}
[ptb]
\begin{center}
\includegraphics[
trim=0.000000in 2.868520in 0.000000in 0.205623in,
height=2.7523in,
width=5.5578in
]%
{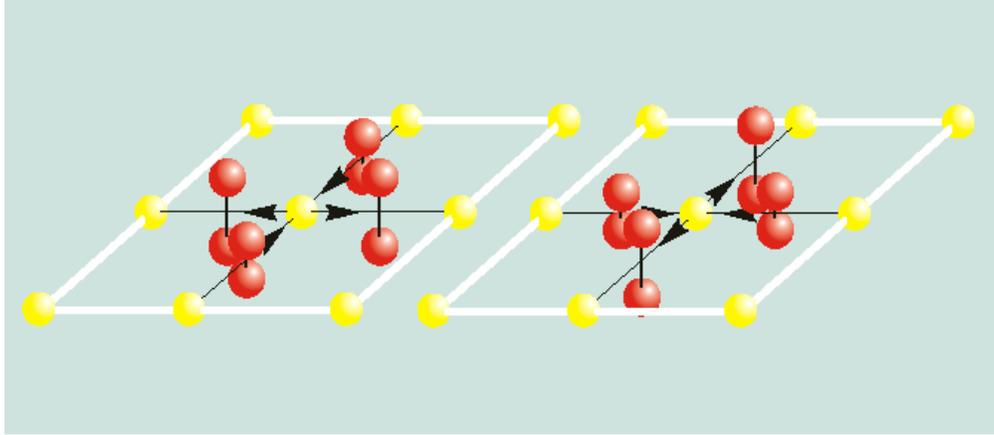}%
\caption{The Fluctuating Bond field after projecting out $s$-fluctuations has
$d$-symmetry, with (left panel) O-amplitude (dumbells) large in $y$-bonds,
small in $x$-bonds, the reverse in right panel. Panels also illustrate
opposite phases of the static $d$-wave CDW (18).}%
\label{Fig4}%
\end{center}
\end{figure}
%EndExpansion

The FBM\ pairing propagator $\mathbf{\Gamma}(\mathbf{k},\mathbf{q},n)$ (Fig.
3) for scattering a pair from $(\mathbf{k,-k})$ to $(\mathbf{k+q,-k-q})$ is
now given (for a full derivation see supplementary material) by the Fig. 3
graphs as a product of $d$-wave form factors and a 2-phonon potential
propagator $\mathbf{V}(\mathbf{q},n)$
\begin{equation}
\mathbf{\Gamma}(\mathbf{k},\mathbf{q},n)=\left\vert \xi_{\mathbf{k,k}%
+\mathbf{q}}\right\vert ^{2}\mathbf{V}(\mathbf{q},n), \label{Gamma}%
\end{equation}
where the $d$-type form factor (coming from the form factor of the bond
operator $X$) is
\begin{equation}
\xi_{\mathbf{k,k}^{\prime}}=\frac{1}{2}\left[  \cos\left(  k_{x}\right)
+\cos\left(  k_{x}^{\prime}\right)  -\cos\left(  k_{y}\right)  -\cos\left(
k_{y}^{\prime}\right)  \right]  ,
\end{equation}
and the potential propagator (dropping a zero-frequency term, only significant
at temperatures above those of interest) is
\begin{equation}
\mathbf{V}(\mathbf{q},n)=\frac{-4n_{s}^{-1}K\omega_{a}^{2}f_{\mathbf{q}}%
}{\omega_{n}^{2}+4\overline{\omega}^{2}+4\omega_{a}^{2}f_{\mathbf{q}}\left[
\frac{1}{2}-KR_{dd}(\mathbf{q},n)\right]  }. \label{V(q,n)}%
\end{equation}
Here $\omega_{n}=2\pi nk_{B}T$ is the Matsubara frequency (Fourier component
with respect to imaginary time), $T=$temperature, $k_{B}=$Boltzmann's
constant, and we introduce definitions of the total vibrator frequency
$\overline{\omega}$, the mean field harmonic vibrator frequency $\omega_{h}$,
and the anharmonic component of the vibrator frequency $\omega_{a}$
\begin{align}
\overline{\omega}^{2}  &  =\omega_{h}^{2}+\omega_{a}^{2},\\
\omega_{h}^{2}  &  =\omega_{0}^{2}+\frac{v}{m}\left\langle X_{\mathbf{i,\alpha
}}\right\rangle ,\nonumber\\
\omega_{a}^{2}  &  =\frac{w}{2m}\left\langle x_{\mathbf{i,\alpha}}%
^{2}\right\rangle =\frac{w}{4m^{2}\overline{\omega}}\coth\left(
\frac{\overline{\omega}}{2k_{B}T}\right)  .\nonumber
\end{align}
In the denominator of (\ref{V(q,n)}) we can identify (in square brackets) the
two terms in the self energy (Fig. 3c), the positive one coming from $w$, and
the negative one coming from $v^{2}R_{dd}$. The other two terms are the
inverse of the two-boson propagator (Fig. 3b) with characteristic frequency
$2\overline{\omega}$ (a frequency up-shifted by the single-boson self energy).
The form factor $f_{\mathbf{q}}$ is defined by $f_{\mathbf{q}}=\frac{1}%
{2}\left[  \cos^{2}\left(  q_{x}/2\right)  +\cos^{2}\left(  q_{y}/2\right)
\right]  ,$ and the $dd$ Response Function $R_{dd}$ is defined in the normal
state as
\begin{equation}
R_{dd}(\mathbf{q},n)=-\sum_{\mathbf{k}}\frac{f(\epsilon_{\mathbf{k}%
})-f(\epsilon_{\mathbf{k+q}})}{i\omega_{n}+\epsilon_{\mathbf{k}}%
-\epsilon_{\mathbf{k+q}}}\xi_{\mathbf{k,k+q}}\xi_{\mathbf{k+q,k}},
\end{equation}
where $f(\epsilon_{\mathbf{k}})$ is the Fermi function. The $dd$ Response
Function (RF) is a generalization of the density-density RF. The presence of
the $\xi$-factors shows that the RF can be interpreted as the $d$-symmetry
density response to a $d$-symmetry perturbation. At long wavelengths, the RF
is decoupled from the on-site fluctuations of s-symmetry, hence it is
decoupled from the strong on-site Coulomb repulsion $U$.

The modified 2-phonon interaction $\mathbf{V}(\mathbf{q},n)$ can diverge at
low frequency and small $q$, at significant values of the interaction $K$, due
to the largeness of $R_{dd}$ when the Fermi level lies at the energy
$\epsilon_{SP}$ of the saddle points (SP) (or "antinodal points")\ at
$\mathbf{k}=\left(  \pi,0\right)  $ and $\left(  0,\pi\right)  $ in the band
structure, signaling the emergence of a 2-phonon bound state. \ The same SP
effect causes a peak, the van Hove singularity, in the density of states (DOS)
at energy $\epsilon_{SP}$. In this situation the superconducting gap acts,
through controlling the magnitude of $R_{dd}$, to regularize the divergence.

Let us now look at the leading-$N$ gap equation at $T_{c}$%
\begin{equation}
\Delta\left(  \mathbf{k,}n\right)  =-T\sum_{\mathbf{k}^{\prime},n^{\prime}}%
\xi_{\mathbf{k,k}^{\prime}}^{2}\mathbf{V}(\mathbf{k-k}^{\prime},n-n^{\prime
})G_{2}\left(  \mathbf{k}^{\prime}\mathbf{,}n^{\prime}\right)  \Delta\left(
\mathbf{k}^{\prime}\mathbf{,}n^{\prime}\right)  ,
\end{equation}
where $\Delta\left(  \mathbf{k,}n\right)  $ is the gap and $G_{2}\left(
\mathbf{k,}n\right)  =\left(  \nu_{n}^{2}+\epsilon_{\mathbf{k}}^{2}\right)
^{-1}$.

In standard BCS theory, $G_{2}$ gives rise to a log divergence in $T$, which
results in a solution to the gap equation with the well-known standard BCS
formula for $T_{c}$ exponential in the inverse coupling constant. In contrast,
in the FBM there is the stronger divergence in $\mathbf{V}(\mathbf{q},m)$,
which tends to peg $T_{c}$ at the temperature where the divergence disappears.
Since this temperature is defined by the denominator in $\mathbf{V}%
(\mathbf{q},m)$, in which no degeneracy factor $n_{s}$ appears, it is a
leading-$N$ formula. Hence we are justified in neglecting the second
Eliashberg equation involving electronic mass renormalization ($Z$-factor) as
a $1/N$ correction.

Solving the gap equation, we always find a $d_{x^{2}-y^{2}}$-wave gap
$\Delta\left(  \mathbf{k,}n\right)  $ due to the $\xi_{\mathbf{k,k}^{\prime}%
}^{2}$ factors - originating in the FBM pairing interaction being localized on
a bond (Fig. 2b) - and to the largeness of the $V(\mathbf{q},n)$ at small $q$.
In Fig. 5a we present some results for $T_{c}$ and oxygen isotope shift (the
technique used was to solve at finite gap using Fast Fourier Transform (FFT)
techniques \cite{Serene}, and extrapolate to zero gap) as a function of
doping. The results show the standard hump in $T_{c}$ as function of doping,
and a very dramatic minimum in the isotope shift going down to almost zero
around the $T_{c}$-maximum, while going up to values above the BCS
$\alpha=0.5$ on the underdoped side. In Fig. 5b we present experimental
results for the "universal" isotope shift behavior found for several materials
\cite{alpha1}, \cite{Tallon} along with the widely used "universal" empirical
formula for $T_{c}$ \cite{formula}. It is seen that there is a remarkable
semiquantitative agreement between theory and experiment, which is especially
remarkable as the isotope shift in BCS theory is quite robust.%
%TCIMACRO{\FRAME{ftbpFU}{6.443in}{3.2251in}{0pt}{\Qcb{(a) Left panel.
%Transition temperature $T_{c}$ and oxygen isotope shift $\alpha^{0}$ vs.
%doping $p$, relative to doping $p_{0}$ at max. $T_{c}$, from gap equation
%(16). Parameters $t=0.25\unit{eV}$, $t^{\prime}=-0.06$ $\unit{eV}$,
%$t^{\prime\prime}=0.0325\unit{eV}$, $K=0.48\unit{eV}$, $\omega_{a}%
%=0.05\unit{eV}$, $\omega_{h}=0.015\unit{eV}$. Phonon frequencies in this range
%are reported by ref's \cite{Opel}-\cite{Saini}(b) Right panel. Experimental
%transition temperature and isotope shift. Full curve, empirical $T_{c}$-doping
%relation $T_{c}=T_{c,\text{\thinspace}\max}(1-82.6\left(  p-p_{0}\right)
%^{2})$ \cite{formula}. Points are oxygen isotope shift measurements for
%YBCO-based and Bi-2212 cuprate superconductors \cite{Tallon}.}}{\Qlb{Fig5}%
%}{fig4abm.eps}{\special{ language "Scientific Word";  type "GRAPHIC";
%width 6.443in;  height 3.2251in;  depth 0pt;  original-width 10.6415in;
%original-height 8.1327in;  cropleft "0.0299";  croptop "1";
%cropright "0.9627";  cropbottom "0.3915";
%filename 'Fig4abm.eps';file-properties "XNPEU";}}}%
%BeginExpansion
\begin{figure}
[ptb]
\begin{center}
\includegraphics[
trim=0.318181in 3.183952in 0.396928in 0.000000in,
height=3.2251in,
width=6.443in
]%
{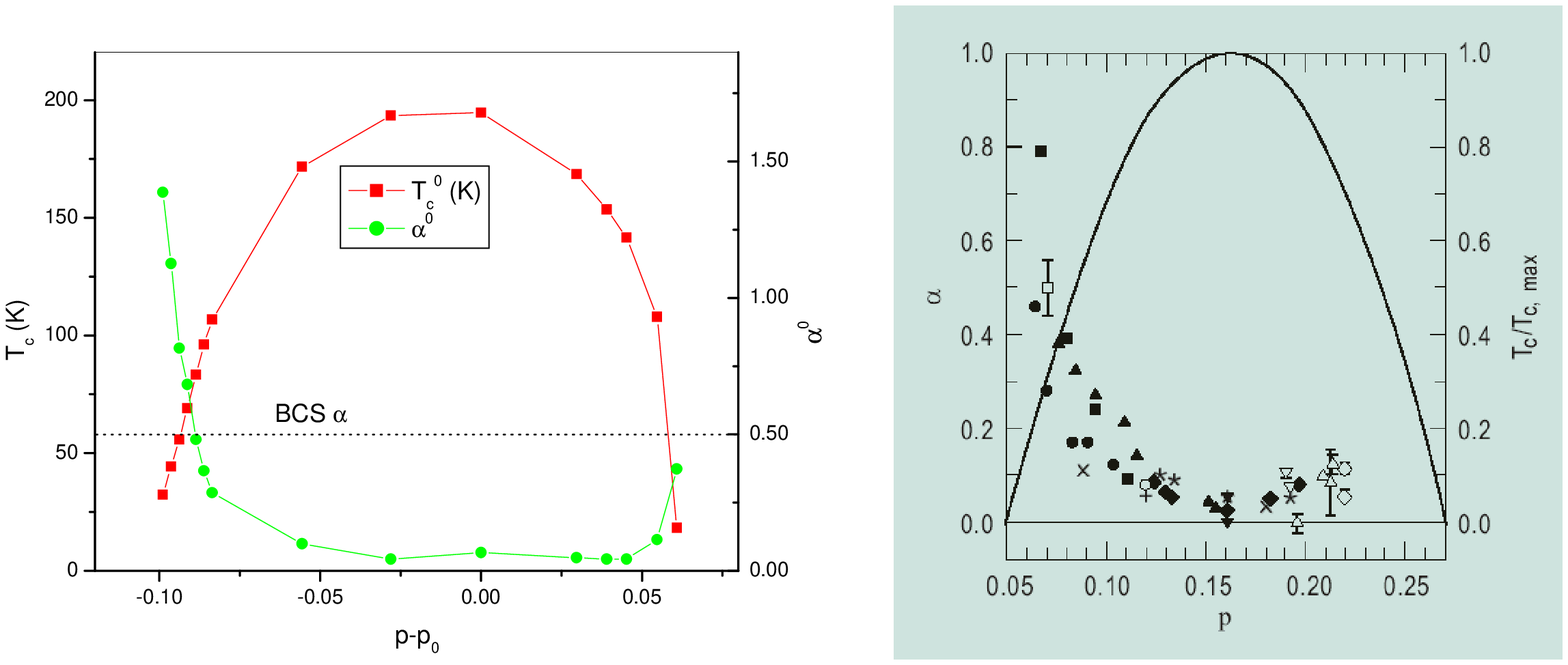}%
\caption{(a) Left panel. Transition temperature $T_{c}$ and oxygen isotope
shift $\alpha^{0}$ vs. doping $p$, relative to doping $p_{0}$ at max. $T_{c}$,
from gap equation (16). Parameters $t=0.25\operatorname{eV}$, $t^{\prime
}=-0.06$ $\operatorname{eV}$, $t^{\prime\prime}=0.0325\operatorname{eV}$,
$K=0.48\operatorname{eV}$, $\omega_{a}=0.05\operatorname{eV}$, $\omega
_{h}=0.015\operatorname{eV}$. Phonon frequencies in this range are reported by
ref's \cite{Opel}-\cite{Saini}(b) Right panel. Experimental transition
temperature and isotope shift. Full curve, empirical $T_{c}$-doping relation
$T_{c}=T_{c,\text{\thinspace}\max}(1-82.6\left(  p-p_{0}\right)  ^{2})$
\cite{formula}. Points are oxygen isotope shift measurements for YBCO-based
and Bi-2212 cuprate superconductors \cite{Tallon}.}%
\label{Fig5}%
\end{center}
\end{figure}
%EndExpansion

The explanation for the hump in $T_{c}$ as a function of doping in the FBM, is
that the density of states and the RF peak around the point where the Fermi
energy coincides with the band structure energy $\epsilon_{SP}$ at the SP. At
this point a zero in the denominator of the pairing interaction $\mathbf{V}%
(\mathbf{q},m)$ is more readily obtained, while high DOS always favors
pairing. The phonon parameters in Fig. 5 have been selected so that the phonon
frequency is mainly of anharmonic origin, i.e. a mainly quartic potential (see
Fig. 2a), when the zero in the denominator of $\mathbf{V}(\mathbf{q},n)$ is
mass-independent ($\omega_{a}$ factors out), hence the extremely low isotope
shift. As the Fermi energy moves away from $\epsilon_{SP}$, the DOS drops and
the FBM divergence in the pairing interaction tends to disappear, with a
resumption of more normal BCS isotope shift.

As well as driving pairing, the coupling $K$ can also produce static
distortions. The 2-phonon interaction $w$ can be renormalized to
$\widetilde{w}$ by summing a somewhat more extended set of leading-$N$ Feynman
diagrams than those in Fig. 2. $\widetilde{w}$ is found to contain the
proportionality factor%
\begin{equation}
\widetilde{w}\sim\frac{1}{2}-KR_{dd}(\mathbf{q},n),
\end{equation}
leading to a zero in the long-wavelength $\widetilde{w}$ at a temperature
$T_{mf}$. Below $T_{mf}$ there is symmetry-breaking in the system, which can
be likened to the presence of an Ising pseodospin in each unit cell. The
symmetry breaking can be described in real space as a local splitting of the
nearest-neighbor hopping integrals $t$, $t_{x}\neq t_{y}$, and in $\mathbf{k}%
$-space as a splitting of the saddle point energies at the $k$-points X and Y,
$\epsilon_{SP}^{X}\neq\epsilon_{SP}^{Y}$.

An ordered pseudospin structure has been sought in the form of a 1D $d$-wave
CDW \cite{Gruner}, with the Ansatz
\begin{equation}
u_{\mathbf{i}}^{2}=\frac{2\sqrt{nn_{s}}}{v}\chi_{\mathbf{Q}}\cos
(\mathbf{Q.i),}%
\end{equation}
where
\begin{equation}
u_{\mathbf{i}}^{2}=\frac{1}{2}\left(  x_{\mathbf{i+}\widehat{\mathbf{x}}%
/2}^{2}+x_{\mathbf{i-}\widehat{\mathbf{x}}/2}^{2}-y_{\mathbf{i+}%
\widehat{\mathbf{y}}/2}^{2}-y_{\mathbf{i-}\widehat{\mathbf{y}}/2}^{2}\right)
, \label{CDW}%
\end{equation}
and $\mathbf{Q}=\left(  Q_{x},Q_{y}\right)  $ is the CDW wavevector. The
$d$-wave nature of the CDW is seen in that the expression (\ref{CDW})
corresponds to a static distortion of the type represented in Fig. 4.

The propagator $G(\mathbf{k},n)$, in a standard approximation where the
self-energy is second order in $\chi_{\mathbf{Q}}$, is given by
\begin{equation}
G(\mathbf{k},n)=\frac{1}{i\nu_{n}-\epsilon_{\mathbf{k}}-\chi_{\mathbf{Q}}%
^{2}\Phi(\mathbf{k},n)},
\end{equation}
with
\begin{equation}
\Phi(\mathbf{k},n)=\frac{1}{4}\left[  \frac{\xi_{\mathbf{k,k+Q}}^{2}}{i\nu
_{n}-\epsilon_{\mathbf{k+Q}}}+\frac{\xi_{\mathbf{k,k-Q}}^{2}}{i\nu
_{n}-\epsilon_{\mathbf{k-Q}}}\right]  .
\end{equation}
By calculating the free energy, the wavevector $\mathbf{Q}$ was determined by
energy minimization. The direction of $\mathbf{Q}$ is found to lie along the
$x$- or $y$- axis, $\mathbf{Q=}\left(  Q,0\right)  $, or $=\left(  0,Q\right)
$. The CDW is found to be stable - relative to the uniformly-polarized (but
presumably disordered) state - below a temperature $T_{CDW}$, which is plotted
in Fig. 6, a stability mostly confined to the underdoped side. $Q$ in this
region closely tracks the nesting wavevector between the two pieces of Fermi
surface at X or Y in the band structure (inset Fig. 6). Note the discrepancy
in the empirical $K$-value between Figures 5 and 6. A large hidden negative
contribution to the empirical $K$ can be assumed from the nearest-neighbor
Coulomb repulsion $V$. In a superconducting context, Coulomb interactions are
reduced to a lower value $V^{\ast}$ by off-shell scattering, while there is no
such effect for the static CDW. Hence the CDW effective $K$ should be less
attractive, with the two $K$-values differing by roughly $V-V^{\ast}$.%
%TCIMACRO{\FRAME{ftbpFU}{6.4524in}{4.4348in}{0pt}{\Qcb{$T_{CDW}$ (blue circles)
%calculated from stability boundary of CDW relative to uniform solution.
%$T_{c}$ (magenta triangles) and $\alpha$ (green triangles) calculated in
%presence of CDW from LG equation. \ Inset shows nesting wavevector
%$\mathbf{Q}$ at point $X=(\pi,0)$ in BZ. Parameters as Fig. 5, except
%$K=0.23\unit{eV}$ in $T_{CDW}$ calculation (see text).}}{\Qlb{Fig6}%
%}{fig_5f.eps}{\special{ language "Scientific Word";  type "GRAPHIC";
%width 6.4524in;  height 4.4348in;  depth 0pt;  original-width 10.6415in;
%original-height 8.1327in;  cropleft "0";  croptop "1";  cropright "1";
%cropbottom "0.1022";  filename 'Fig_5f.eps';file-properties "XNPEU";}}}%
%BeginExpansion
\begin{figure}
[ptb]
\begin{center}
\includegraphics[
trim=0.000000in 0.831162in 0.000000in 0.000000in,
height=4.4348in,
width=6.4524in
]%
{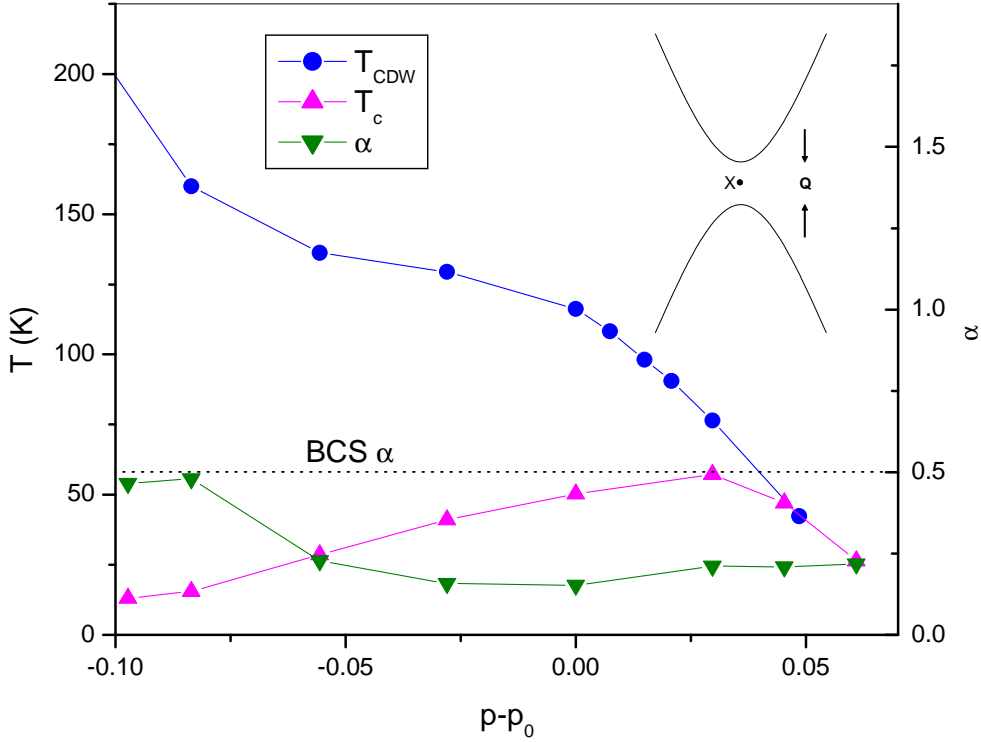}%
\caption{$T_{CDW}$ (blue circles) calculated from stability boundary of CDW
relative to uniform solution. $T_{c}$ (magenta triangles) and $\alpha$ (green
triangles) calculated in presence of CDW from LG equation. \ Inset shows
nesting wavevector $\mathbf{Q}$ at point $X=(\pi,0)$ in BZ. Parameters as Fig.
5, except $K=0.23\operatorname{eV}$ in $T_{CDW}$ calculation (see text).}%
\label{Fig6}%
\end{center}
\end{figure}
%EndExpansion

The CDW-induced modification to the energy band structure can be seen by
looking at the poles in the propagator $G(\mathbf{k},n)$ lying at low energy.
Suppose that $\epsilon_{\mathbf{k}}$ and $\epsilon_{\mathbf{k+Q}}$ lie near
the Fermi level ($\mathbf{k}$ and $\mathbf{k+Q}$ are 2 points related by
nesting), when $\epsilon_{\mathbf{k-Q}}$ will typically lie far away so only
the term in $\epsilon_{\mathbf{k+Q}}$ in $\Phi$ diverges. Then the poles are
approximately at
\begin{equation}
i\nu_{n}=\epsilon_{F}\pm\frac{1}{2}\chi_{\mathbf{Q}}\xi_{\mathbf{k,k+Q}%
}\rightarrow\epsilon_{F}\pm\chi_{\mathbf{Q}},
\end{equation}
where $\xi_{\mathbf{k,k+Q}}\simeq\pm2$. An identical argument can be made for
the case of scattering from $\mathbf{k}$ to $\mathbf{k-Q}$. Hence we see that
$\chi_{\mathbf{Q}}$ is the gap opened up by the breakdown in long range order
caused by the CDW - it is a form of pseudogap. The factor $\xi_{\mathbf{k,k+Q}%
}$, which is approximately $\cos k_{x}-\cos k_{y}$ at the small $Q$ of
interest, shows the presence of a $d$-wave symmetry factor in this gap.
Because of nesting factors, the gap is not symmetric in X vs. Y, but full
symmetry in $k$-space will reappear on averaging over spatial domains
$\mathbf{Q=}\left(  Q,0\right)  $, or $=\left(  0,Q\right)  $.

We tentatively identify the CDW gap as the $d$-wave pseudogap seen in the HTS
materials, and identify $T_{CDW}$ with $T^{\ast}$. Experimental studies show a
highly inhomogeneous spatial variation of the gap \cite{McElroy}, apparently
controlled by the oxygen dopant distribution, which we interpret as a strongly
impurity-pinned CDW. The regions where the gap is maximal ($75$ meV - our
$\chi_{\mathbf{Q}}\simeq60$ meV), but with little sign of superconducting
coherence peaks, are interpreted as the peaks in the CDW amplitude. The length
scale $\simeq40%
%TCIMACRO{\unit{\U{212b}}}%
%BeginExpansion
\operatorname{\text{\AA}}%
%EndExpansion
$ observed is similar to that predicted from $Q$.

We need to estimate the interaction between the two order parameters,
superconducting and CDW. We calculate the CDW\ order parameter on the basis of
the normal state, assuming that always $T_{CDW}>$ $T_{c}$ ($T^{\ast}>T_{c}$),
for which there is supporting evidence \cite{Lia}. The superconducting phase -
due to its small coherence length $\xi\sim2%
%TCIMACRO{\unit{nm}}%
%BeginExpansion
\operatorname{nm}%
%EndExpansion
$ - can coexist with the competing CDW phase \cite{Matsubara}. We shall here
describe the inhomogeneous two-order parameter coexistence phase in an
approximate manner using a form of Landau-Ginzburg (LG) approach. The approach
does not take into account local distortion of the superconducting order
parameter by the CDW (local distortions average out, leaving perfect $d$-wave,
over a CDW wavelength).

The LG expression for the superconducting free energy takes the form%

\begin{equation}
F=\frac{a}{2}\left(  \nabla\Delta\right)  ^{2}+\frac{b}{2}\left(  T-T_{c}%
^{0}\right)  \Delta^{2}+\frac{1}{2}\Delta^{2}f\left(  \chi^{2}\left(
\mathbf{x}\right)  \right)  ,
\end{equation}
where $\chi\left(  \mathbf{x}\right)  $ is the CDW order parameter
\begin{equation}
\chi\left(  \mathbf{x}\right)  =\chi_{\mathbf{Q}}\cos\left(  \mathbf{Q.x}%
\right)  ;\;Q\neq0,\text{ \ \ }\chi=\chi_{\mathbf{0}}/\sqrt{2};\;Q=0,
\end{equation}
$\Delta(\mathbf{x})$ is the superconducting gap, $a$ and $b$ and are LG
parameters, $T_{c}^{0}$ is the transition temperature in the absence of the
CDW order parameter, and $f$ is a coupling function between the two order parameters.

In the absence of CDW\ order the coherence length is given by the standard
formula $\xi_{LG}^{0}=\sqrt{a/bT_{c}^{0}}$ $=0.739\xi_{BCS}^{0}$; the BCS
coherence length $\xi_{BCS}^{0}$ is assumed to follow BCS scaling relative to
the known coherence length of the HTS. The function $f$ is determined from
solving the gap equation as a function of a spatially uniform CDW amplitude
$\chi_{\mathbf{0}}$, the input being the splitting $t_{x}-t_{y}=\chi
_{\mathbf{0}}$ in nearest-neighbor hopping integral. To limit order parameter
distortion, a cutoff on the maximum $\chi$ is inserted, which however does not
affect the conclusions.

Differentiating the free energy w.r.t. $\Delta$ we get the Schrodinger-like
equation (SE)\ for the gap%
\begin{equation}
-a\nabla^{2}\Delta+V(\mathbf{x})\Delta=\epsilon\Delta,
\end{equation}
where $V(\mathbf{x})=f\left(  \chi^{2}\left(  \mathbf{x}\right)  \right)
-bT_{c}^{0}$, $\epsilon=-bT$. The lowest energy $\epsilon$ solution represents
the highest global transition temperature $T_{c}=T$. The procedure is then to
(1) solve for the CDW amplitude and wavevector, (2) calculate the effect on
$T_{c}$ of a given magnitude of uniform order parameter, and then (3) solve
the SE\ for the transition temperature.

The results are shown in Fig. 6. We see that there is a significant reduction
in $T_{c}$ coming from the effect of the CDW. The isotope shift still has the
same qualitative behavior, but with less variation - because $\chi
_{\mathbf{0}}$ provides an additional perturbation competing with the effect
of chemical potential. The STM results\cite{McElroy} might be more compatible
with a 2D CDW than the 1D CDW assumed in this paper, and indeed tentative
results for a 2D CDW show much less $T_{c}$-reduction, and enhanced isotope
shift variation, but in the absence of pinning the 1D CDW is found to be more
stable than the 2D one.

In the strong coupling limit, where the effect of $K$ is larger than the
effect of $t$, preliminary studies suggest that the FBM can produce a striped
phase and a transition to an insulating state. Combined with $U$, a true Mott
transition involving magnetism can result, leading to an RVB-like state.

We believe that this work constitutes a breakthrough in finding a natural
phonon-based mechanism capable of generating a $d$-wave superconducting gap, a
$d$-wave pseudogap, and giving the type of behavior of $T_{c}$, $T^{\ast}$,
and isotope shift similar to that observed experimentally. The expected
inhomogeneity of the order parameter is also observed \cite{McElroy}.
Quantitative comparisons should bear in mind that the current picture is a
highly simplified single band, single interaction parameter one. A future
program of work comprises a rather long list, including first principles
substantiation of the anharmonic oxygen potential and coupling to the
electronic degrees of freedom, inclusion of the oxygen 2$p$-bands, calculation
of spectroscopic properties such as the superconductivity-induced shift in
vibrator frequency, and the phonon peaks seen in tunneling spectroscopy of the
superconducting state, together with understanding the fundamental energy
balance - possibly unconventional - underlying HTS. \ A better understanding
of the CDW order and its interaction with superconductivity should be obtained
from a more detailed coherent, quantum-mechanical, theoretical approach.

\begin{acknowledgments}
Both authors contributed equally to this work. Correspondence should be
addressed to D.M.N at dennisn@us.ibm.com
\end{acknowledgments}

\end{document}